\documentclass[12pt]{article}
\pdfoutput=1
\usepackage[english]{babel}
\usepackage{dirtytalk}
\usepackage[utf8x]{inputenc}
\usepackage{multirow}
\usepackage[english]{isodate}
\title{Bibliography management: \texttt{natbib} package}
\usepackage{arxiv}
\usepackage{tablefootnote}
\usepackage{apacite}
\usepackage{filecontents}

\usepackage{amssymb,amsmath,amsfonts,eurosym,geometry,ulem,graphicx,caption,color,setspace,sectsty,comment,natbib, footmisc,caption,pdflscape,subfigure,array,hyperref}
\usepackage{lipsum}
\newcolumntype{L}[1]{>{\raggedright\let\newline\\arraybackslash\hspace{0pt}}m{#1}}
\newcolumntype{C}[1]{>{\centering\let\newline\\arraybackslash\hspace{0pt}}m{#1}}
\newcolumntype{R}[1]{>{\raggedleft\let\newline\\arraybackslash\hspace{0pt}}m{#1}}

\begin{document}

\title{Of Access and Inclusivity
Digital Divide in Online Education}

\author{
 Bheemeshwar Reddy A \\
 Assistant Professor \\
  Department of Economics and Finance \\
  Birla Institute of Technology and Science Pilani, Hyderabad \\
   \And
 Sunny Jose \\
 RBI Chair Professor\\
  Council for Social Development \\
  Southern Regional Centre, Hyderabad \\
  \And
 Vaidehi R \\
 Doctoral Scholar \\
    Department of Economics and Finance \\
    Birla Institute of Technology and Science Pilani, Hyderabad \\
}

\printdate{2020-08-31}
\maketitle
Can online education enable all students to participate in and benefit from it equally? Massive online education without addressing the huge access gap and disparities in digital infrastructure would not only exclude a vast majority of students from learning opportunities but also exacerbate the existing socio-economic disparities in educational opportunities.
\section{Introduction}
The Covid-19 pandemic has thrown the educational system of the country in disarray. The closure of schools and colleges began before the completion of the end-semester or annual exams and cast a blight on the entire academic cycle. A reinforcing web of issues, such as prolonged closure, uncertainty about the timing of reopening, likely constriction in the academic calendar and the resultant learning discontinuity among students, among others, has forced the states and educational institutions to find a feasible option to assuage the varied impacts. Online education has emerged as a preferred, predominant option. In fact, it has been pontificated to the status of TINA—there is no alternative to it. On the face of it, online education appears as a safe interim bet. Yet, its scale, scope and reach raise serious issues, both on the process preceding it and the outcome proceeding from it.
\par
In the process, the state and educational administrations, behaving as “benevolent patriarchs,” have taken this key decision without the participation of important stakeholders in the educational system, namely teachers, students and parents. Also, the decision of massive online education is rationalised as a Hobson’s choice of either online education or no education at all. Furthermore, cloaked behind is the assumption that online teaching is structurally no different from in-person teaching and education is primarily content delivery (\cite{Brabazon2020}). Whether teachers are willing, suitably equipped with necessary skills and are comfortable in teaching classes online are questions snubbed as irrelevant at the least or obstructive at the worst.
\par
Most importantly, the decision to launch massive online education neglects a crucial factor on which it is critically contingent: students’ access to digital infrastructure, namely a mix of either computer, tablet or smartphone and access to the internet. The neglect of this vital input raises serious questions on the outcome aspect. Do the households in general and those who have school/ college going students in particular have access to essential digital infrastructure? Will online education enable all students to participate in and profit from it equally? Or, will it leave behind those who lack access to the digital infrastructure? This commentary engages with these issues.
\par
Analysis of the National Sample Survey Office (NSSO) data on social consumption of education (2017–18) informs that only about 9\% of students who are currently enrolled in any course have access to essential digital infrastructure, and such measly access is enmeshed with huge socio-economic and spatial disparities. Hence, the attempt to make online education an opportunity out of the Covid-19 crisis poses a serious risk of leaving many students, especially the socio-economically disadvantaged, further behind (\cite{UNESCO2020}).
\par
\begin{figure}[h]
\caption{Percentage of Households with Access to Digital Infrastructure in India}
\centering
\includegraphics{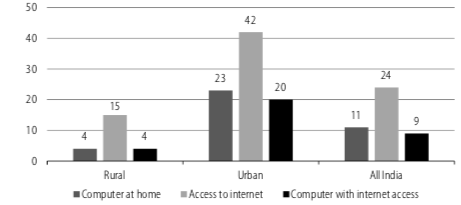}
{\raggedright  \centering Source: Estimated from NSSO data on social consumption of education (2017–18).\par}
\end{figure}

\section{Access to Digital Infrastructure}
As part of the survey on social consumption of education (75th round, held in 2017–18), the NSSO collected information regarding having a computer as well as access to and use of the internet at home, among others. We examine the access to digital infrastructure by analysing this data. The survey considers a household having a computer if it has in its possession any device, such as desktop computer, laptop computer, notebook, netbook, palmtop, tablet (or similar handheld devices). Access to the internet is defined as a household possessing any device, such as computer, tablet, smart phone, personal digital assistant, game machine, digital television, etc, through which internet facility is available, whether or not it has been used. The data, however, does not specify which device was used to access internet. This caveat needs to be kept in mind while discussing access to internet.
\par
About 24\% of households in India had access to the internet through any of the digital devices, whereas only 11\% had computer (including tablet), the most suitable device to partake in online education, in 2017–18 (Figure 1). However, the proportion slides sharply when these two aspects are combined: computer with internet access. The spatial disparities in these aspects are quite pronounced. Only a paltry 4\% of rural households have access to computer with internet, the corresponding proportion is 20\% for urban households—about five times the access in rural India. However, this also implies that over 80\% of households in urban India do not have access to computer with internet. What is more shocking is that while over 42\% of urban households have access to internet through any of the digital devices, including smart phones, it is only 15\% in rural India. This low access to internet militates against the modern myth that India today is empowered with the magic wand of extensive internet access.
\par
\begin{figure}[h]
\caption{Percentage of Currently Enrolled Students with Access to Digital Infrastructure}
\centering
\includegraphics{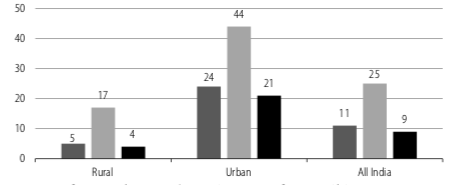}
{\raggedright  \centering Source: Same as Figure 1 \par}
\end{figure}

\par
Specifically, we need to look at students with access to digital infrastructure (5–35 years) who are currently enrolled in any course, as online education is mainly meant for them. Here too, about 9\% of the currently enrolled students have access to computer with internet, implying that over 91\% of students are left behind from having access (Figure 2).
The proportion of students having access is higher in urban (21\%) than rural (4\%) India. What about access to internet among the currently enrolled students? Only 25\% of them have access to internet through any of the devices. The access is much better in urban India (44\%) than rural India (17\%). Can massive online education be inclusive when 75\% and 91\% of the currently enrolled students do not have access to internet and computer with internet, respectively? It appears that the problem of exclusion might surpass the benefit of inclusion.

\begin{table}[]
\centering
\caption{Percentage of Currently Enrolled Students with Access to Digital Infrastructure in States}
\begin{tabular}{llllllllll}
\hline
\multirow{2}{*}{\textbf{State}} & \multicolumn{3}{c}{\textbf{Rural}}                   & \multicolumn{3}{c}{\textbf{Urban}}                   & \multicolumn{3}{c}{\textbf{All India}}               \\ \cline{2-10} 
                                & \textbf{CwI} & \textbf{Computer} & \textbf{Internet} & \textbf{CwI} & \textbf{Computer} & \textbf{Internet} & \textbf{CwI} & \textbf{Computer} & \textbf{Internet} \\ \hline
All-India                       & 4            & 5                 & 17                & 21           & 24                & 44                & 9            & 11                & 25                \\
Andhra   Pradesh                & 2            & 2                 & 13                & 9            & 13                & 31                & 4            & 6                 & 20                \\
Arunachal   Pradesh             & 5            & 10                & 10                & 26           & 32                & 34                & 8            & 13                & 14                \\
Assam                           & 4            & 5                 & 13                & 30           & 34                & 45                & 7            & 8                 & 17                \\
Bihar                           & 3            & 3                 & 15                & 18           & 20                & 40                & 4            & 5                 & 18                \\
Goa                             & 27           & 35                & 69                & 31           & 34                & 75                & 30           & 34                & 73                \\
Gujarat                         & 4            & 4                 & 22                & 21           & 24                & 52                & 10           & 12                & 34                \\
Haryana                         & 6            & 7                 & 40                & 28           & 31                & 58                & 13           & 15                & 46                \\
Himachal   Pradesh              & 11           & 13                & 58                & 28           & 30                & 78                & 13           & 14                & 60                \\
Jammu   and Kashmir             & 5            & 6                 & 34                & 16           & 17                & 65                & 7            & 8                 & 41                \\
Jharkhand                       & 1            & 2                 & 14                & 13           & 14                & 45                & 4            & 4                 & 21                \\
Karnataka                       & 1            & 1                 & 9                 & 16           & 20                & 27                & 6            & 8                 & 15                \\
Kerala                          & 19           & 23                & 53                & 24           & 27                & 62                & 21           & 24                & 57                \\
Madhya   Pradesh                & 2            & 2                 & 11                & 14           & 17                & 37                & 5            & 6                 & 18                \\
Maharashtra                     & 3            & 4                 & 22                & 25           & 28                & 54                & 12           & 14                & 36                \\
Manipur                         & 7            & 9                 & 34                & 18           & 18                & 56                & 11           & 13                & 42                \\
Meghalaya                       & 3            & 5                 & 8                 & 24           & 30                & 44                & 7            & 9                 & 15                \\
Mizoram                         & 12           & 16                & 35                & 34           & 47                & 59                & 23           & 31                & 46                \\
Nagaland                        & 17           & 18                & 40                & 44           & 51                & 71                & 25           & 28                & 50                \\
Odisha                          & 1            & 2                 & 7                 & 15           & 18                & 28                & 4            & 5                 & 10                \\
Punjab                          & 6            & 8                 & 41                & 30           & 32                & 63                & 14           & 16                & 48                \\
Rajasthan                       & 7            & 9                 & 22                & 26           & 28                & 50                & 12           & 13                & 28                \\
Sikkim                          & 16           & 18                & 75                & 35           & 35                & 74                & 21           & 22                & 75                \\
Tamil   Nadu                    & 9            & 13                & 17                & 20           & 26                & 29                & 14           & 19                & 23                \\
Telangana                       & 1            & 2                 & 12                & 16           & 19                & 41                & 8            & 10                & 26                \\
Tripura                         & 1            & 1                 & 7                 & 9            & 11                & 23                & 3            & 3                 & 10                \\
Uttar   Pradesh                 & 4            & 5                 & 14                & 19           & 21                & 40                & 7            & 8                 & 19                \\
Uttarakhand                     & 5            & 7                 & 37                & 30           & 34                & 67                & 12           & 14                & 44                \\
West   Bengal                   & 3            & 3                 & 9                 & 22           & 23                & 39                & 8            & 9                 & 17               
\end{tabular}
{\raggedright  \centering Computer with internet access is abbreviated as CwI \par Source: Same as Figure 1.}
\end{table}

Across the states, Goa emerges as the top performer with 30\% of currently enrolled students having access to computer with internet (Table 1, p 24), followed by Nagaland (25\%), Mizoram (23\%), Kerala and Sikkim (21\%). At the bottom of the access ladder are Tripura (3\%), Jharkhand (4\%), Andhra Pradesh, Bihar, Odisha (4\%) and Madhya Pradesh (5\%). Even in richer states (Gujarat, Maharashtra and Punjab) access varies from 10\% to 14\%. Recall here that access to computer includes tablet.
\par
In rural parts of the states, access to computer with internet is a meagre 1\% in five states, such as Jharkhand, Karnataka, Odisha, Telangana and Tripuraa whopping 99\% of currently enrolled students lack access to this vital mix. The access is equally low (2\% to 4\%) in many bigger states (Andhra Pradesh, Madhya Pradesh, West Bengal, Bihar, Maharashtra, Gujarat and Uttar Pradesh). Only in six states that remain at the top end, the access is more than 10\%: Himachal Pradesh (11\%), Mizoram (12\%), Sikkim (16\%), Nagaland (17\%), Kerala (19\%) and Goa (27\%). In urban parts, the currently enrolled students have relatively better access, though access is not very high. Nagaland with 44\% access emerges as the top per- former, followed by Sikkim (35\%). In five states (Assam, Punjab, Uttarakhand, Goa and Mizoram) the access is above 30\%. Andhra Pradesh and Tripura with only 9\% access emerge as the poor performers. Jharkhand, Madhya Pradesh and Odisha follow them closely.
\par
In access to internet among currently enrolled students, Sikkim (75\%) and Goa (73\%) emerge as top performers. By contrast, Tripura and Odisha with only 10\% of access remain at the bottom. In 16 states spread across various regions of India, less than one-third of the currently enrolled students are lucky enough to have access to internet. These include some of the bigger states as well (Bihar, Madhya Pradesh, Uttar Pradesh, West Bengal, Andhra Pradesh, Telangana, Tamil Nadu and Rajasthan). Only in five states, the access is more than 50\%.
\par
The access to internet is less than 10\% in rural parts of Odisha, Tripura, Meghalaya, Karnataka and West Bengal. By contrast, only in four states (Sikkim, Goa, Himachal and Kerala) the access is more than 50\%. In 18 states, the access is less than 25\%. In urban parts, more than 70\% of the currently enrolled students have access in Himachal Pradesh, Sikkim, Goa and Nagaland (states with higher access), whereas only less than one-third of the students have access to internet in five states (Tripura, Karnataka, Odisha, Tamil Nadu and Andhra Pradesh), the states with lower access.
\par
Two major points emerge from the above discussion. First, the overall access to digital infrastructure essential for online education is paltry in rural areas of most of the states. Though access is higher in urban parts, there too around 80\% of the currently enrolled students do not have access to computer with internet. Second, the lower access to digital infrastructure is not confined to fewer states belonging to specific regions or with higher poverty. Rather, these states by defying any specific categorisation indicate the peculiar tenacity of poor availability of digital infrastructure in general. This has serious implications for the effectiveness of online education, which is being deployed as a feasible, effective alternative to in-person teaching.
\begin{table}[]
\centering
\caption{Percentage of Students with Access to Digital Infrastructure in Socio-religious and Income Groups}
\begin{tabular}{llll}
\hline
\textbf{Groups}         & \textbf{Computer With Internet} & \textbf{Computer} & \textbf{Internet} \\ \hline
Scheduled Tribes        & 4                               & 5                 & 13                \\
Scheduled Castes        & 4                               & 5                 & 17                \\
Other Backward Classes  & 7                               & 9                 & 23                \\
All   Muslims           & 8                               & 9                 & 23                \\
Others                  & 21                              & 23                & 45                \\ \hline
\textbf{Income deciles} & \textbf{}                       & \textbf{}         & \textbf{}         \\ \hline
1                       & 2                               & 3                 & 10                \\
2                       & 2                               & 2                 & 9                 \\
3                       & 3                               & 3                 & 12                \\
4                       & 3                               & 4                 & 16                \\
5                       & 4                               & 5                 & 19                \\
6                       & 4                               & 5                 & 21                \\
7                       & 7                               & 8                 & 26                \\
8                       & 10                              & 12                & 33                \\
9                       & 16                              & 19                & 44                \\
10                      & 41                              & 45                & 66       
\end{tabular}
\par
    {\raggedright  \centering Source: Same as Figure 1.\par}
\end{table}
\section{Digital Divide}
It appears that the currently enrolled students from advantaged socio-economic groups have relatively better access to digital infrastructure (Table 2). The access to computer with internet is the highest among students from the top income decile (richest 10\%). Also, 66\% and 45\% of them have access to computer at home and internet, respectively. Among the socio-religious groups, students from the advantaged social group have the highest access. Between advantaged social group (others) and top income group (richest 10\%), the access is almost double among the latter in computer with internet and computer at home. This suggests that economic advantage seems to outweigh social advantage in augmenting access to digital infrastructure.
\par
Students from the rest of the groups have low to moderate access. The access to computer with internet is only 8\% among students from the Other Backward Classes and 10\% among students from the eighth decile (third richest group). Higher the socio-economic disadvantage, much lesser the access. Only 2\% of students from the poorest income groups have access to computer with internet, only 3\% have access to computer at home and 10\% have access to internet through any of the digital devices. Students from Scheduled Tribes (STs) and Scheduled Castes (SCs) also have an equally measly access. Here too, material deprivation seems to matter more than social marginality in limiting access to digital infrastructure.
\par
Our estimates of concentration index intending to identify the extent of inequality in access to digital infrastructure between income groups suggests that inequality in access to computer with internet is the lowest in Goa, Sikkim, Tamil Nadu and Kerala (results not shown here). By contrast, inequality is the highest in Karnataka, Tripura, Odisha, West Bengal and Madhya Pradesh. The case of Goa, Sikkim and Kerala appears interesting, as in these states not only is inequality in access to digital infrastructure the lowest, but also the overall access to digital infrastructure is much higher. The reverse holds good in Tripura, Odisha, West Bengal and Madhya Pradesh.
\section{Conclusion}
Online education is considered the best possible opportunity out of the Covid-19 crisis to palliate its impacts on the education sector. Many states and educational institutions have already embraced on- line education as a primary means of teaching for the next academic year or semester. It is against this context we examined the access to digital infrastructure among currently enrolled students and the extent of digital divide between students from socio-economic groups in India.
\par
The analysis brings out at least three important facts. First, a vast majority of currently enrolled students did not have access to essential digital infrastructure, computer or tablet with internet access in India in 2017–18. The proportion is as high as 90\%. To put it differently, only 9\% of students were fortunate enough to have access to the suitable digital infra- structure. In rural India, only a “blessed few students” (4\%) had access to this crucial infrastructure. Though access is relatively better in urban India, here too around 80\% of students lack access to computer with internet. It appears that access to digital infrastructure among students has become an exception than the rule. The situation improves margin- ally in access to internet. Only 25\% of the currently enrolled students had access to internet through any of the digital devices, including smart phone in India in 2017–18.
\par
Second, across the states the access to computer with internet is the highest in Goa with 30\%. The access is quite low in rural parts of many states. Though access goes up in urban parts of the states, it is not high either. The states where access is low are many, spread across all regions of India. Defying any categorisation, they point to the generalised phenomenon of limited access to digital infrastructure in India. Third, the access to essential digital infrastructure is woefully low among students from the poorest income and social groups. Only students from the richest income group and advantaged social group have higher access.
\par
The above facts raise a number of important questions. How inclusive and enabling would be online education in India when 90\% of the currently enrolled students do not have access to digital infrastructure? What sort of inclusion and learning, online education can possibly foster when just 4\% of students from STs and SCs have access? What would happen to 98\% of students from the poorest income groups? The double whammy of low access and deep digital divide will possibly exclude a large majority of students from actively participating in and benefiting from online education. Hence, launching of massive online education without sufficiently ad- dressing both, the huge access gap and disparities in digital infrastructure would lead to the exclusion of socio-economically disadvantaged students from learning opportunities. Most importantly, such an exclusion will possibly exacerbate the already existing huge and systemic socio- economic disparities in educational opportunities and outcomes.


\bibliography{EPW}  

\begin{thebibliography}{2}
\providecommand{\natexlab}[1]{#1}
\providecommand{\url}[1]{\texttt{#1}}
\expandafter\ifx\csname urlstyle\endcsname\relax
  \providecommand{\doi}[1]{doi: #1}\else
  \providecommand{\doi}{doi: \begingroup \urlstyle{rm}\Url}\fi

\bibitem[Brabazon and Honor(2020)]{Brabazon2020}
Brabazon and Honor.
\newblock The academy’s neo-liberal response to covid-19: Why faculty should
  be wary and how we can push back?
\newblock \emph{Academic Matters}, 2020.

\bibitem[UNESCO(2020)]{UNESCO2020}
UNESCO.
\newblock Global education monitoring report 2020: Inclusion and education: All
  means all, 2020.

\end{thebibliography}







\newcommand\blfootnote[1]{%
  \begingroup
  \renewcommand\thefootnote{}\footnote{#1}%
  \addtocounter{footnote}{-1}%
  \endgroup
}
\blfootnote{This is an Accepted Manuscript version of the following article, accepted for publication in Economic and Political Weekly.}
\end{document}